# Textbook myths about early atomic models

Reidun Renstrøm, Nils-Erik Bomark*

University of Agder, Kristiansand, Norway

Abstract

Most physics textbooks at college and university level introduce quantum physics in a historical context. However, the textbook version of this history does not match the actual history. In this article, the first in a series of articles looking at the textbook description of the quantum history, we follow an exceptional student through her endeavors to understand the early atomic models that led up to the work of Niels Bohr. We experience her disappointment when she discovers that the description of the famous atomic model by Thomson, is a mere caricature with almost no trace of Thomson's work and that the supposedly important radiative instability of Rutherford's atoms, is not there at all; how could it be? Rutherford did not even discuss the motion of the electrons. These flaws in the narrative are not just cosmetic inaccuracies, the main narrative is centered around claims that are simply wrong or non-existent in the real story and thus rendering the whole telling mere fiction. We as a physics community should do better.

*nilseb@uia.no

# 1. Introduction

Almost all physics textbooks introduce atomic physics in a historical context. The story is framed historically, giving the impression that the description is a summarized account of the course of events that led Bohr to his atomic model and a quantum theory of line spectra (Renstrøm,2011).

It begins with a presentation of Joseph John Thomson's atomic model from 1904, the most popular model of the atom until 1910. Most textbooks depict the model as a sphere of positive charge embedded with electrons, like a cake filled with raisins. According to the textbook story, this model could explain the atom's emission spectra. In 1910-11, Ernest Rutherford carried out alpha scattering experiments designed to test Thomson's atomic model. The results could only be explained with a positively charged nucleus in the center of the atom, and therefore, the story goes, Rutherford's model had the electrons orbiting the nucleus as planets revolving around the sun. Thus, Thomson's model had to be taken offstage. However, the textbooks explain, orbiting electrons would lose energy through electromagnetic radiation and hence spiral into the nucleus while the frequency of revolution increases. Therefore, the atoms should collapse, but in reality, they remain stable for an infinite period of time. Faced with this grave anomaly, the young Danish physicist Niels Bohr decided to go to England to work together with Rutherford to understand how solar-system-like atoms could be stable and not radiate away all their energy.

The problem is that this textbook or standard story is closer to a fairytale than to true history. This is unfortunately a common problem when the history of discoveries in physics is presented, many of the famous tales all physics students are exposed to are mere quasi-stories.

The prehistory and development of Niels Bohr's quantum theory of line spectra has the potential to improve students' understanding of quantum concepts, promote their understanding of the nature of physics and to give them a good opportunity to discover physics as a human activity. The obvious condition for these benefits is that the history that the students learn is historically accountable and not a distorted version that supports a certain view of how new knowledge progresses in physics. Today, young people learn to trust the science, but to be wary of fake news and distorted history. When students discover that the history presented in physics textbooks is largely a made-up story, they will undoubtedly be very disappointed. It was not what they had expected.

This paper is the first of a series of papers, exposing the flaws in the textbook version of the historical development of quantum theory. Here we will follow the unusually dedicated and curious student Emmy as she reads about early atomic models and how she time and again discovers that the stories in her textbooks do not match what she finds when reading the original papers.

After, in section 2, listing the books used in this study, we start with atomic model by J.J. Thomson in section 3, followed by a brief note about Nagaoka's model in section 4. In section 5 we look at Rutherford and his exploration of atomic physics and a brief note about Nicholson's model in section 6 and finally the conclusions in section 7.

## 2. Textbooks used in this study

To examine the presentations of the prehistory of Bohr's atomic model and a quantum theory of line spectra in some present-day physics textbooks, we have chosen textbooks published during the period 2009–2014. These contemporary books are in relatively widespread use in college and university physics departments in several countries.

The textbooks examined for the purposes of the present paper about Thomson's model are:

- *University Physics, 13th ed.* (Young & Freedman, 2012)
- *Physics for Scientists and Engineers with Modern Physics, 9th ed.* (Jewett & Serway, 2014)

In addition to these textbooks we have examined a popular book "Thirty years that shook physics", a book about the development of modern physics published in 1966 written by the famous physicist George Gamow. Gamow was a student of Niels Bohr and during his years at Niels Bohr's Institute he had met many of the scientists who contributed to the early development of quantum theory. Therefore, the history Gamow tells appears particularly trustworthy, and his book has been an inspiration for many professional physicists and physics students for many years.

- *Thirty years that shook physics. The Story of Quantum Theory* (Gamow, 1966)

# 3. Thomson's atomic model

The first atomic model Emmy encounters is the model by J.J. Thomson, this model was introduced by the discoverer of the electron in a seminal paper in 1904 and was the most popular model of the atom between 1904 and 1910.

Emmy excitedly reads what her books says about this model.

## 3.1 The textbook version of Thomson's model

All her books describe the model in a similar way; there is a uniform spherical positive charge with electrons scattered (seemingly randomly) throughout it, as illustrated in the figure 1. It is often compared to raisins in a cake or a plum-pudding.

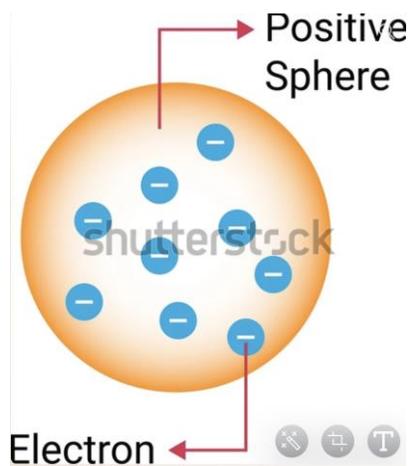

*Figure 1. The common depiction of Thomson´s model with electrons randomly distributed in a positive charge sphere, as raisins in a cake.*

In the words of *University Physics* section 39.2:

> In 1910 the best available model of atomic structure was one developed by Thomson. He envisioned the atom as a sphere of some as yet unidentified positively charged substance, within which the electrons were embedded like raisins in cake.

Physics for Scientists and Engineers with Modern Physics puts it like this.   (p.1299 ):

> The following year, he suggested a model that describes the atom as a region in which positive charge is spread out in space with electrons embedded throughout the region, much like the seeds in a watermelon or raisins in thick pudding (Fig. 42.3). The atom as a whole would then be electrically neutral.

Thirty years that shook physics, says that (p.29):

> Thomson visualized the atom as being formed by some positively substance distributed more or less uniformly through the entire body, with negatively charged electrons imbedded in it as are raisins inside a round loaf of raisin bread.

Emmy feels that this description sounds a bit too simplistic. A model in physics must allow us to do calculations. This sounds like some loose speculations at a conference dinner, not a full-fledged atomic model by a Nobel laureate physicist.

She also wonders why this model became so popular. University Physics offers one explanation:

> This model offered an explanation for line spectra. If the atom collided with another atom, as in a heated gas, each electron would oscillate around its equilibrium position with a characteristic frequency and emit electromagnetic radiation with that frequency. If the atom were illuminated with light of many frequencies, each electron would selectively absorb only light whose frequency matched the electron's natural oscillation frequency.

However, Thirty years that shook physics, does not agree that this is possible:

> If Thomson´s model of an atom was accepted, it was possible, by the methods of classical mechanics, to calculate the equilibrium distribution of electrons within the body of an atom containing a given amount of inner electrons, and it was expected that the sets of calculated characteristic vibration frequencies would coincide with the observed line spectra of various elements. Thomson himself and his students carried out complicated computations to find the configurations of the interatomic electrons for which the calculated frequencies in the line spectra of various chemical elements. The results were disappointingly negative. The theoretically calculated spectra based on Thomson´s model looked not at all like the observed spectra of any of the chemical elements.

So why did this model become so popular? Emmy searches the internet and soon she finds the original article by Thomson from 1904 published in *Philosophical Magazine* in Martz in 1904 (Thomson, 1904).

### 3.2. Thomson's paper "On the structure of the atom"

The title of the paper is:

> *On the structure of the atom: an investigation of the stability and periods of oscillation of a number of corpuscles arranged at equal intervals **around the circumference of a circle**; with application of the results to the theory of atomic structure* [our emphasis]

The paper's title is enough for Emmy to realize that the corpuscles (Thomson's name for electrons) are not embedded in Thomson's atom model like raisins in a cake. They are arranged in circles, and the paper investigates the stability of these circles.

In the first sentence of the paper's introduction, Thomson presents a picture of his model of the structure of the atom:

> *"The view that atoms of the elements consist of a number of negatively electrified corpuscles enclosed in a sphere of uniform positively electrification, suggests, among other interesting mathematical problems, the one that discussed in this paper, that of the motion of the ring of* n *negatively electrified particles placed inside a uniformly electrified sphere"* (p. 237)

The configuration of electrons in Thomson's model is a solution to interesting but complicated mathematical problems. For example, an atom with 60 electrons has five rings with 3, 8, 13, 16 and 20 electrons in each. Thomson's calculations of the motions of the electrons arranged in rings did not depend on the mass of the positive charge, which were therefore irrelevant. Furthermore, he showed that if one electron was displaced from the equilibrium configuration, it would be driven back to it. Thomson mastered what few thought possible; namely, to create a stable mechanical model of the atom. The only function of the positive charge is to keep the electrons together.

In Thomson's own words:

> *In this way we see that when we have a large number of corpuscles in rapid rotation, they will arrange themselves as follows: - The corpuscles form a series of rings, the corpuscles in one ring being approximately in a plane at right angles to the axis of rotation, the number of particles in the rings diminishing as the radius of the ring diminishes. If the corpuscles can move at right angles to the plane of their orbit, the rings will be in different planes adjusting themselves so that the repulsion between the rings is balanced by the attraction exerted by the positive electrification of the sphere in which they are placed … The number of corpuscles in a ring vary from ring to ring; each corpuscle is travelling at a high speed around the circumference of the ring in which it is situated, and the rings are so arranged that those which contain a large number of corpuscles are near to the surface of the sphere, while those in which there are a smaller number of corpuscles are more in the inside.* (p. 254)

Again, Emmy wonders why physicists insist on describing this as raisins in a cake when the electrons are in fact arranged in carefully designed circles that rapidly rotate around the central axis of the atom. This does not look like any plum pudding she ever heard of!

Thomson used his model first to explain qualitatively the periodic table of the elements and the nature of radioactivity and then other phenomena as chemical combination and the scattering of $\beta$ —rays. Thomson's model could also in a qualitative manner, explain the emission of light from atoms as a result of the vibrations of electrons, as long as the number of electrons was of the same order as the number of spectral lines. But the model could not explain the regularities known from line spectra, such as Balmer's formula (Kragh, 1999), (Mehra & Rechenherg, 1982).

During the next few years, Thomson conducted many experiments to decide the number of electrons in atoms. In 1906 he concluded that there are few electrons in the atom– the number of electrons is not in the thousands, but comparable with the atomic weight (Thomson, 1906).

In Thomson's own words:

> *I consider in this paper three methods of determining the number of corpuscles in an atom of an elementary substance, all of which lead to the conclusion that this number is of the same order as the atomic weight of the substance. The data at present available indicate that the number of corpuscles in the atom is equal to the atomic weight. As, however, the evidence is rather indirect and the data are not very numerous, further investigation is necessary before we can be sure of this quality; the evidence at present available seems, however, sufficient to establish the conclusion that the number of corpuscles is not greatly different from the atomic weight.* (p. 169)

In 1909, Max Born praised Thomson's model as "being like a piano excerpt from the great symphonies of luminating atoms" (Kragh,1999 p.46). Emmy agrees that this is an impressive model, a classical model of the atom that is mechanically stable. However, she is befuddled by the descriptions she found in her textbooks that seem to picture something else entirely. Why do physicists call it the plum-pudding model if it does not look like a plum-pudding? Why do the pictures of the model look nothing like the model Thomson described? Are physicists not supposed to be very careful in their presentations so that what they say is accurate and defendable?

## 4. Nagaoka's Saturnian model

In her studies of Thomson's model, Emmy discovers that there were more models of the atom around at the time. For instance, also published in *Philosophical Magazine* in Martz in 1904. Hantora Nagaoka, a professor of physics at the Imperial University of Japan, also presented an atomic model (Nagaoka, 1904).

This model resembled a Saturnian atom with very many electrons moving in one or more rings around a positively charged central body. In Nagaoka's own words:

> *The system, which I am going to discuss, consists of a large number of particles of equal mass arranged in a circle at equal angular intervals and repelling each other with forces inversely proportional to the square of distance; at the center of the circle, place a particle of large mass attracting the other particles according to the same law of force.*
>
> *If these repelling particles be revolving with nearly the same velocity about the attracting center, the system will generally remain stable, for small disturbances provided the attracting force be sufficiently great…the present case will evidently be approximately realized if we replace these satellites by negative electrons and the attracting center by a positively charged particle.*

A serious problem of the Saturnian model was a lack of stability for the electron orbits. The comparison with Saturn fails because of the repulsion between electrons and the model I mechanically unstable (Pais, 1986 p. 183). This accentuates that Thomson's model was the most popular exactly because of its mechanical stability.

## 5. Rutherford's atomic model

Few events in the history of physics are more famous than Rutherford's experiment, and every student of physics is introduced to it as a landmark in the history of the discipline, an epochal turning point in the history of physics.

### 5.1 The textbook version of Rutherford's findings

According to all of Emmy's books the Rutherford experiment took Thomson's model off the stage and revealed that the real atom has a positively charged nucleus with electrons circling around it, like planets around the Sun.

But more importantly, physicists in 1911 had to realize that classical physics failed to explain the stability of Rutherford's experimentally based atomic model. Maxwell's theory of electromagnetism predicted a catastrophe: atoms are unstable and will emit radiation continuously, as illustrated in figure 2. A crisis had entered physics.  The only way out of a crisis like this is to develop a theory that can explain the experimental facts.

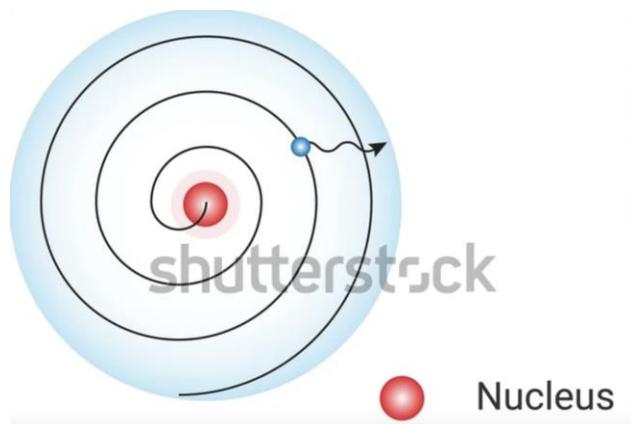

*Figure 2. The figure illustrates that the size of the electrons orbit decreases until they fall into the nucleus because the acceleration of the electrons makes them radiate away their energy.*

University Physics: 39.2 Rutherford's Exploration of the Atom, p.1294:

> The first experiments designed to test Thomson's model by probing the interior structure of the atom were carried out in 1910-1911 by Ernest Rutherford (Fig 39.10) and two of his students, Hans Geiger and Ernest Marsden, at the University of Manchester in England. These experiments consisted of shooting a beam of charged particles at thin foils of various elements and observing how the foil deflected the particles… … In the Thomson model, the positive charge and the negative electrons are distributed through the whole atom. Hence the electric field inside the atom should be quite small, and the electric force on an alpha particle that enters the atom should be quite weak. The maximum deflection to be expected is then only a few degrees
>
> The results of the Rutherford experiments were very different from the Thomson prediction. Some alpha particles were scattered by nearly 180°— that is, almost straight backward (Fig. 39.12b). Rutherford later wrote:
>
> **It was quite the most incredible event that ever happened to me in my life. It was almost as incredible as if you had fired a 15-inch shell at a piece of tissue paper and it came back to you**.
>
> Clearly the Thomson model was wrong and a new model was needed. Suppose the positive charge, instead of being distributed through a sphere with atomic dimensions (of the order of $10^{-10}$ m), is all concentrated in a much smaller volume. Then it would act like a point charge down to much smaller distances. The maximum electric field repelling the alpha particle would be much larger, and the amazing large angle scattering that Rutherford observed could occur. Rutherford developed this model and called the concentration of positive charge the nucleus. He again computed the numbers of particles expected to be scattered through various angles. Within the accuracy of his experiments, the computed and measured results agreed, down to distances of the order of $10^{-14}$ m. His experiments therefore established that the atom does have a nucleus—a very small, very dense structure, no longer than $10^{-14}$ m in diameter)

The other books tell a very similar story and the quote from Rutherford appears more than once. From here the story goes on to tell us about "The Failure of Classical Physics", again from University Physics:

> Rutherford's discovery of the atomic nucleus raised a serious question. What kept the negatively charged electrons at relatively large distances away from the very small, positively charged nucleus despite their electrostatic attraction? Rutherford suggested that perhaps the electrons revolve in orbits around the nucleus, just as the planets revolve around the sun. But according to classical electromagnetic theory, any accelerating charge radiates electromagnetic waves … The energy of an orbiting electron should therefore

> decrease continuously, its orbit should become smaller and smaller, and it should spiral rapidly into its nucleus (Fig. 39.14).
>
> Even worse, according to classical theory the frequency of the electromagnetic waves emitted should be equal to the frequency of revolution. As the electrons radiated energy, their angular speeds would change continuously, and they would emit a continuous spectrum (a mixture of all frequencies) not the line spectrum actually observed. Thus Rutherford's model of electrons orbiting the nucleus, which is based on Newtonian mechanics and classical electromagnetic theory, make three entirely wrong predictions about atoms: They should emit light continuously, they should be unstable, and the light they emit should have a continuous spectrum. Clearly a radical reappraisal of physics on the scale of the atom was needed.

In Physics for Scientists and Engineers with Modern Physics, p.1299 we read:

> Rutherford explained his astonishing results by developing a new atomic model, one that assumed the positive charge in the atom was concentrated on a region that was small relative to the size of the atom. He called this concentration of positive charge the nucleus of the atom. Any electrons belonging to the atom were assumed to be in the relatively large volume outside the nucleus. To explain why these electrons were not pulled into the nucleus by the attractive electric force, **Rutherford modeled them as moving in orbits around the nucleus in the same manner as the planets orbit the Sun**
>
> For this reason, this model is often referred to as the planetary model of the atom. Two basic difficulties exist with Rutherford's planetary model. … As energy leaves the system, the radius of the electron's orbit steadily decreases (. … Therefore, as the electron moves closer to the nucleus, the angular speed of the electron will increase, just like the spinning skater in Figure 11.10 in Section 11.4. This process leads to an ever-increasing frequency of emitted radiation and an ultimate collapse of the atom as the electron plunges into the nucleus.

Thirty years that shook physics puts it like this (Gamow, 1966, pp. 35-36):

> Thus was born Rutherford´s atomic model. With its light negatively charged electrons moving through free space around a positively charged heavy nucleus in the center, it somewhat resembled the Solar System. ….atomic electrons move around the nucleus along the circular or elliptic orbits, just as do planets around the Sun. …It is well known that the oscillating electric charges produce diverging electromagnetic waves….Using the classical theory of electromagnetic emission, one easily calculates that light waves emitted by electrons circling the atomic nucleus will take away into space all the electron´s energy within one hundred-millionth of a second. Having lost all their energy, atomic electrons must fall into the nucleus and the atom cease to exist!

Emmy finds this story somewhat odd; why is it such a big problem that Rutherford puts electrons in circles when every single atomic model she heard about does the same? And what about Nagaoka's model? It looks very similar to Rutherford's, and nobody seemed bothered with the radiative stability of that; the mechanical instability was an issue but no radiative instability was mentioned (Heillbron & Kuhn, 1969)

Again, a quick search on the internet brings forth the original paper by Rutherford from 1911 (Rutherford, 1911)

## 5.2. Rutherford's paper
The title of the paper is:

*The scattering of α and β Particles by Matter and the Structure of the Atom*

Rutherford writes:

> *Since the α and β particles traverse the atom, it should be possible from a close study of the nature of the deflexion to form some idea of the constitution of the atom to produce the effects observed. In fact, the scattering of high-speed charged particles by the atom of matter is one of the most promising methods of attack of this problem. We shall first examine theoretically the single encounters with an atom of simple structure, which is able to produce large deflections of an α particle, and then compare the deductions from the theory with the experimental data available.* **Consider an atom which contains a charge ±Ne at its center surrounded by a sphere of electrification containing a charge ∓Ne supposed uniformly distributed throughout a sphere of radius R.** *e is the fundamental unit of charge, which in this paper is taken as 4.65 x $10^{-10}$ E.S. unit. … It will be shown that the main deductions from the theory are independent of whether the central charge is supposed to be positive or negative. For convenience, the sign will be assumed to be positive. The question of the stability of the atom proposed need not be considered at this stage, for this will obviously depend upon the minute structure of the atom, and on the motion of the constituent charged parts.* (Rutherford, 1911, pp. 670-671)[our emphasis]

Rutherford did not suggest a planetary atom in 1911. He offered no suggestions of how the electrons were arranged or moved; the electrons' exact distribution was not important for the experimental results. The theory Rutherford presented in his paper was primarily a scattering theory, and it was *not* considered to be an atomic model. Rutherford's model was met with indifference and scarcely considered to be a theory of the constitution of the atom. It was not mentioned in the proceedings of the 1911 Solvay Congress, nor was it widely discussed in physics journals. Not even Rutherford himself seemed to have considered the nuclear atom to be of great importance. In his textbook *Substances and their Radiations* from 1913, only 7 pages of 700 dealt with the new discovery. It was a lack of interest in the nuclear atom, first because he didn´t offer **any suggestion** of how the electrons were arranged. The negative electricity forms a homogenous atmosphere around the nucleus. The electrons were of no importance in the scattering (Kragh, 1999).

In his paper, *The structure of the atom* (Rutherford, 1914), Rutherford repeated the conclusion he drew about the structure of the atom in 1911:

> *In order to account for this large scattering of particles, I supposed that the atom consisted of a positively charged nucleus of small dimensions in which practically all the mass of the atom was concentrated.* **The nucleus was supposed to be surrounded by a distribution of electrons to make the atom electrically neutral** *and extending to distance from the nuclear comparable with the ordinary accepted radius of the atom.* (pp. 488-489) [our emphasis]

A model of this sort cannot be in static equilibrium since the electrons would all be dragged into the nucleus. Rutherford was aware of this, but he chose to ignore the difficulty for the time being. He stated that the question of the stability of the atom proposed need not be considered at this stage…

Emmy is now rather disappointed in her textbooks; why is every single book talking so much about the radiative instability of electrons in circular orbits, when Rutherford never even mentions how the electrons move? As a matter of fact, he is the only one who does not put the electrons in circles!

She also discovers that Thomson in 1903 demonstrate that radiative instability is not much of a problem as long as the number of electrons is not very small (Thomson, 1903); for larger number of electrons the circular motion looks more like a continuous current and hence only produce static electromagnetic fields. That seems to be why people did not worry so much about radiative stability for the early models.

## 6. Nicholson's atomic model

Before the conclusion we would like to mention the model by John Nicholson (Nicholson, 1912). This model is usually not mentioned in textbooks but did play a crucial role in the development of our understanding of the atom. It is in fact the first model where the angular momentum of the electrons can only change discontinuously so in some sense it is quantized.

This model was developed between 1911 and 1914 and it is somewhat similar to Nagaoka's. Nicholson assumed that all atoms consist of a combination of four primary atoms, which existed only in the stars. The primary atoms had a positive charged nucleus in the center of rotating electrons. As Nagaoka and Thomson, Nicholson explained the emitted light of a certain frequency by vibrations of the electron rings with same frequency. In the third paper from June 1912 he concluded that in the primary atoms the angular moment $L$ of the electron rings had values that were integer $n$ multiples of Planck's constant divide by $2\pi$;

$$L = n\frac{h}{2\pi}.$$

Nicholson's concluded that the angular moment could only rise or fall by discrete amounts; it was quantized. Nicholson's model was successful to explain lines in the spectra of stellar nebulae and the solar corona, but not the Balmer spectrum of hydrogen. This model included Planck's constant in a classical description of the emission and absorption of radiation.

Common for the atomic model architects Thomson, Nagaoka and Nicholson are that they let the electrons move in rings and they explain emitted radiation by vibration of electrons, as required by classical physics. They are all radiatively unstable, but in addition, Nagaoka´s and Nicholson´s model is also mechanically unstable. Rutherford is the only atom-architect who did not pay any attention to the electrons' movement.

## 7. Conclusions

The history of the development of our atomic knowledge is an interesting one, however, as Emmy has demonstrated, the common story told in the textbook falls apart very fast when looked upon more closely. The standard depiction of Thomson's atomic model as a "plum-pudding" bears no resemblance with Thomson's actual work. The catastrophic radiative instability of Rutherford's model does not exist there since he does not talk about the electrons. Discovering these blatant misrepresentations risk significantly reduce Emmy's trust in the whole field of physics.

Perhaps someone might say that these are just a few errors in the story. But that is certainly not the case, the main points of the story are simply wrong which means the entire story is fiction.

Of course, few if any students do have the curiosity and competence to pursue these questions as thoroughly as Emmy, so teachers will usually get away with telling the textbook story, but we as a community should not be satisfied with this situation!

Physics is perhaps the most exact of all sciences and we pride ourselves with extremely well tested and precise laws that allows us to make firm statements about the world we live in. This should be reflected in the way we present our field to the outside world as well.

**References**

Gamov, G., (1966), *Thirty Years That Shook Physics. The Story of Quantum Theory*. Garden City, New York: Anchor Books Doubleday & Company, Inc.

Heilbron, J.L. & Kuhn, T., (1969), The genesis of the Bohr atom. *Historical Studies in the Physical Sciences,* Vol 1, 211-290) University of California Press

Jewett J.W and Serway R.A, (2014), *Physics for Scientists and Engineers with Modern Physics* , 9th ed. 2014, Boston: Brooks/Cole.

Kragh, H., (1999), *Quantum Generations: A history of Physics in the Twentieth Century*. Princeton, New Jersey: Princeton University Press.

Mehra, J., & Rechenherg, H., (1982), *The Historical Development of Quantum Theory* (Vol. 1). New York: Springer-Verlag.

Nagaoka, H., (1904), Kinetics of a system of particles illustrating the line and the band spectra and the phenomena of radioactivity. *Philosophical Magazine, 7*, 445-455.

Nicholson, J. W., (1912), The constitution of the solar corona. III. *Monthly Notices Roy. Astr. Soc.* (London) *72,* 677-692

Pais, A., (1986), *Inward Bound*: Oxford University Press.

Renstrøm, R., (2011), *Kvantefysikkens utvikling - i fysikklærebøker, vitenkapshistorien og undervisning [The development of quantum physics - in physics textbooks, in the history of science, and in the classroom].* (Ph.D. thesis), University of Oslo

Rutherford, E. J., (1911), The Scattering of α and β Particles by Matter and The Structure of the Atom. *Philosophical Magazine, 21*, 669-688.

Rutherford, E. J., (1914), The structure of the atom. *Philosophical Magazine, 27*, 488-498

Thomson, J. J., (1903), The magnetic properties of systems of corpuscles describing circular orbits , *Philosophical Magazine 6*, 673-693.

Thomson, J. J., (1904), On the structure of the atom. *Philosophical Magazine, 7*, 237-265.

Thomson, J. J., (1906), On the number of corpuscles in an atom, *Philosophical Magazine* 6, 769-781.

Young, H.D. and R.A. Freedman, (2012), *University Physics* 13th ed. 2012, San Fransisco: Pearson Addison Wesley.